\documentclass[prl,aps,twocolumn,superscriptaddress,preprintnumbers,10pt]{revtex4-1}
\usepackage{ifpdf}
\ifpdf
\usepackage{graphicx,color}
\usepackage{hyperref}
\else
\usepackage[dvipdfmx]{graphicx,color}
\usepackage[dvipdfmx]{hyperref}
\fi
\usepackage{bm}
\usepackage{amsmath,amssymb,amsfonts}
\usepackage{url}
\usepackage{soul}

\usepackage[capitalise]{cleveref}

\newcommand{\Ocal}[0]{\mathcal{O}}
\newcommand{\MPl}[0]{M_\mathrm{Pl}}
\newcommand{\VGUT}[0]{\mathrm{VGUT}}
\newcommand{\DGUT}[0]{\mathrm{DGUT}}
\newcommand{\ovl}{\overline}
\newcommand{\tr}[0]{\mathrm{tr}}
\newcommand{\ep}{\epsilon}
\newcommand{\lam}{\lambda}
\newcommand{\Si}{\Sigma}
\newcommand{\eqs}[1]{\begin{equation}\begin{split} #1 \end{split}\end{equation}}

\pdfoutput=1
\begin{document}

\preprint{IPMU19-0094}
\preprint{CTPU-PTC-19-19}

\title{
Baryon-Dark Matter Coincidence in Mirrored Unification
}

\author{Masahiro Ibe}
\affiliation{
Kavli IPMU (WPI), UTIAS, The University of Tokyo, Kashiwa, Chiba 277-8583, Japan
}
\affiliation{
ICRR, The University of Tokyo, Kashiwa, Chiba 277-8582, Japan
}
\author{Ayuki Kamada}
\affiliation{
Center for Theoretical Physics of the Universe,
Institute for Basic Science (IBS), Daejeon 34126, Korea
}
\author{Shin Kobayashi}
\affiliation{
Kavli IPMU (WPI), UTIAS, The University of Tokyo, Kashiwa, Chiba 277-8583, Japan
}
\affiliation{
ICRR, The University of Tokyo, Kashiwa, Chiba 277-8582, Japan
}
\author{Takumi Kuwahara}
\affiliation{
Center for Theoretical Physics of the Universe,
Institute for Basic Science (IBS), Daejeon 34126, Korea
}
\author{Wakutaka Nakano}
\affiliation{
Kavli IPMU (WPI), UTIAS, The University of Tokyo, Kashiwa, Chiba 277-8583, Japan
}
\affiliation{
ICRR, The University of Tokyo, Kashiwa, Chiba 277-8582, Japan
}

\date{\today}

\begin{abstract}
About 80\% of the mass of the present Universe is made up of the unknown (dark matter), while the rest is made up of ordinary matter.
It is a very intriguing question why the {\it mass} densities of dark matter and ordinary matter (mainly baryons) are close to each other.
It may be hinting the identity of dark matter and furthermore structure of a dark sector.
A mirrored world provides a natural explanation to this puzzle.
On the other hand, if mirror-symmetry breaking scale is low, it tends to cause cosmological problems.
In this letter, we propose a mirrored unification framework, which breaks mirror-symmetry at the grand unified scale, but still addresses the puzzle.
The dark matter mass is strongly related with the dynamical scale of QCD, which explains the closeness of the dark matter and baryon masses.
Intermediate-energy portal interactions share the generated asymmetry between the visible and dark sectors.
Furthermore, our framework is safe from cosmological issues by providing low-energy portal interactions to release the superfluous entropy of the dark sector into the visible sector.
\end{abstract}

\maketitle

\section{Introduction}
Cosmological observations have established that the mass of the present Universe is made up by so-called dark matter (DM) in addition to ordinary matter.
The mass density of DM is about five times larger than that of ordinary matter, i.e., standard model (SM) baryons~\cite{Aghanim:2018eyx}.
The observed closeness of the mass densities may be a hint on DM and dark sector physics.
If DM (dark sector) has nothing to do with SM baryon (visible sector), it is puzzling why their mass densities are close to each other.

The concept of a mirror world is a natural option to explain this puzzle (see Refs.~\cite{Kobzarev:1966qya,Blinnikov:1983gh,Kolb:1985bf,Khlopov:1989fj,Hodges:1993yb,Foot:2004pa,Chacko:2005pe} for earlier works).
In recent years, the mirror world scenarios combined with twin Higgs models also attract attention since they ameliorate the naturalness problem~\cite{Farina:2015uea,Craig:2015xla,Garcia:2015toa,Barbieri:2016zxn}.
In those scenarios, the dark sector contains mirror partners of the SM particles, and therefore the coincidence is naturally realized.
However, if mirror $\mathbb{Z}_2$ symmetry is kept at a low-energy scale, mirror-world models tend to be inconsistent with cosmology because the dark sector inevitably includes light particles such as the mirror partners of neutrinos and photon.

Instead, in this letter, we pursue a mirrored Grand Unified Theory (GUT) framework, in which $\mathbb{Z}_2$ symmetry is broken at a GUT scale.
We consider a GUT model with gauge dynamics of $G_\VGUT \times G_\DGUT$  (with a gauge group $G = G_\VGUT = G_\DGUT$) and an exchanging symmetry between $G_\VGUT$ and $G_\DGUT$~\footnote{
Refs.~\cite{Lonsdale:2014wwa,Gu:2014nga} are in a similar line but with focusing on the coincidence of the dynamical scales~\cite{Lonsdale:2014wwa} and with low-energy $\mathbb{Z}_2$ breaking~\cite{Gu:2014nga}.
}.
It is remarkable that $\mathbb{Z}_2$-symmetry breaking at a high-energy scale does not lose good features as long as the lightest ``dark'' baryons are DM~\cite{Gudnason:2006yj, Dietrich:2006cm, Khlopov:2007ic, Khlopov:2008ty, Foadi:2008qv, Mardon:2009gw, Kribs:2009fy, Barbieri:2010mn, Blennow:2010qp, Lewis:2011zb, Appelquist:2013ms, Hietanen:2013fya, Cline:2013zca, Appelquist:2014jch, Hietanen:2014xca, Krnjaic:2014xza, Detmold:2014qqa, Detmold:2014kba, Asano:2014wra, Brod:2014loa, Antipin:2014qva, Hardy:2014mqa, Appelquist:2015yfa, Appelquist:2015zfa, Antipin:2015xia, Hardy:2015boa, Co:2016akw, Dienes:2016vei, Ishida:2016fbp, Lonsdale:2017mzg, Berryman:2017twh, Mitridate:2017oky, Gresham:2017zqi, Gresham:2017cvl, Ibe:2018juk, Gresham:2018anj, Francis:2018xjd, Braaten:2018xuw, Bai:2018dxf, Chu:2018faw} (see Ref.~\cite{Kribs:2016cew} for a review).
The baryon-DM coincidence puzzle is divided into two subproblems: the coincidence of masses and that of number densities between baryons and DM.
As for the mass coincidence, the key ingredient is the correspondence between dynamical scales of each sector: the baryon and DM masses are determined by them.
Such a correspondence can be achieved once the gauge couplings are related with each other at the GUT scale. 
As we will see, $\mathbb{Z}_2$-symmetry breaking below the GUT scale does not spoil this correspondence.

As for the number density, we consider the asymmetric dark matter (ADM) framework~\cite{Nussinov:1985xr,Barr:1990ca,Barr:1991qn,Dodelson:1991iv,Kaplan:1991ah,Kuzmin:1996he,Fujii:2002aj,Kitano:2004sv,Gudnason:2006ug,Kitano:2008tk,Kaplan:2009ag}
(see also Refs.~\cite{Davoudiasl:2012uw,Petraki:2013wwa,Zurek:2013wia} for reviews).
Since the ``dark'' baryons have an annihilation cross section as large as the SM baryons have~\cite{Alves:2009nf,Alves:2010dd}, the number density of DM is dominated by an asymmetry between particle and antiparticle.
The asymmetries in the two sectors are equilibrated when a portal interaction is efficient at an intermediate scale.

After decoupling of the portal interaction, the entropy densities in the two sectors are conserved separately: the excessive entropy in the dark sector gives a significant contribution to the dark radiation~\cite{Blennow:2012de}.
Therefore, two types of portal interactions are needed for viable (composite) ADM scenarios: intermediate-energy portal interactions to share the asymmetry, and low-energy portal interactions to release the superfluous entropy of the dark sector into the visible sector.
Our framework indeed provides such portal interactions, and thus explains the baryon-DM coincidence puzzle in a self-contained manner.

Our framework is based on a supersymmetric Grand Unified Theory (SUSY GUT), in which the $\mathbb{Z}_2$ symmetry is manifest above the GUT scale.
The gauge structure of each sector at low energy depends on a choice of vacuum at GUT scale.
In our framework, the visible sector is reduced to the SM, while the dark sector follows two-step symmetry breaking and then has a dynamics similar to quantum chromodynamics (QCD) and quantum electrodynamics (QED).
The second symmetry breaking in the dark sector provides the intermediate-energy portal interactions and tiny kinetic mixing of visible photon and ``dark'' photon.
SUSY plays a key role to achieve gauge coupling unification in the visible sector. 
Electroweak symmetry breaking in the visible sector and ``dark'' QED breaking are triggered by SUSY breaking effects.

\section{Mirrored Unification Model}

We consider a concrete model with $G = SU(5)$ to demonstrate our framework.
$\mathbb{Z}_2$ is the symmetry interchanging $SU(5)_\VGUT$ and $SU(5)_\DGUT$.
Under the $\mathbb{Z}_2$ symmetry, dimensionless couplings in the two sectors are identified, while the mass parameters softly break the $\mathbb{Z}_2$ symmetry.

\begin{table}[t]
	\centering
	\caption{
	Matter and Higgs contents in the $SU(5)_\VGUT \times SU(5)_\DGUT$ model.
	The subscript $S=V, D$ represents the sectors: the fields are charged under $SU(5)_\VGUT$ for $S=V$ and charged under $SU(5)_\DGUT$ for $S = D$.
	$i = 1, 2, 3$ denotes the generations of matter chiral multiplets in each sector.
	\label{tab:contents}
	}
	\begin{tabular}{|c|c|c|}
		\hline
		& $SU(5)_{\VGUT,\DGUT}$ & $U(1)_X$ \\ \hline
		$\Psi_{Si}$ & $\mathbf{10}$  & $1$ \\
		$\Phi_{Si}$ & $\ovl{\mathbf{5}}$  & $-3$ \\ \hline
		\hline
		$\ovl N_i \,, \ovl N'_i$ & $\mathbf{1}$  & $5$ \\ \hline
		\hline
		$H_S$ & $\mathbf{5}$  & $-2$ \\
		$\ovl H_S$ & $\ovl{\mathbf{5}}$  & $2$ \\
		$X_S$ & $\mathbf{5}$  & $-2$ \\
		$\ovl X_S$ & $\ovl{\mathbf{5}}$  & $2$ \\
		$\Si_S$ & $\mathbf{24}$ & $0$ \\
		\hline
	\end{tabular}
\end{table}

We show the particle contents of the model in \cref{tab:contents}, which are similar to those of the minimal SUSY $SU(5)$ GUT in each sector.
The chiral multiplets, $\Psi_V$ and $\Phi_V$, contain all the SM fermions.
$\Psi_D$ and $\Phi_D$ include the dark-quarks which provide the ingredients of the composite DM.
$\ovl N_i$ and $\ovl N'_i$ are the right-handed neutrinos, which are doublets under the ${\mathbb Z}_2$ symmetry.
$U(1)_X$ denotes a global $B-L$ symmetry compatible with the unified gauge group.
It should be noted that the model have extra Higgs quintuplets, $(X_S,\bar{X}_S)$, in addition to the usual Higgs quintuplets, $(H_S,\bar{H}_S)$.

The minimal $SU(5)$ GUT model with $\Ocal(1)\,\mathrm{TeV}$ sfermions contradicts with the nucleon decay experiments~\cite{Goto:1998qg,Murayama:2001ur,Abe:2014mwa,Takhistov:2016eqm}.
To avoid rapid nucleon decay, we simply assume a split spectrum for sparticles~\cite{Giudice:2004tc,ArkaniHamed:2004fb,Wells:2004di,Ibe:2006de}, where sfermions have masses of $\Ocal(10^2)\,\mathrm{TeV}$ while the masses of gauginos and higgsinos are $\Ocal(1)\,\mathrm{TeV}$~%
\footnote{
Thanks to the supersymmetric counterpart of the kinetic mixing parameter $\ep \simeq 10^{-9}$, our framework is free from the late-time decay of the heavier LSP to the lighter one when the LSPs in two sectors are higgsinos~\cite{Ibe:2018tex}.
}.

Both sectors are mostly sequestered with each other up to higher-dimensional interactions
suppressed by the reduced Planck mass $\MPl$.
The superpotential $W_S$ gives the Yukawa couplings, the Higgs masses, and the Higgs couplings to fields with subscripts $S = V, D$,
\eqs{
W_S & = \Psi_S Y_u \Psi_S H_{S}
+ \Psi_S Y_d \Phi_S \ovl H_{S} \\
& ~ + H_S (M_S + \lam \Si_S) \ovl H_S \\
& ~ + \mu_S \tr(\Si_S^2) + \lam_\Si \tr(\Si_S^3) \\
& ~ + M_S' X_S\ovl X_S - \xi\displaystyle{\frac{(X_S \ovl X_S)^2}{\MPl}}
\,.
\label{eq:sp}
}
Here, $\lam$, $\lam_\Sigma$, $\xi$, and $3\times 3$ matrices $Y_{u,d}$ are dimensionless coupling constants, while $M_S$, $M_S'$ and $\mu_S$ are dimensionful parameters.
We assume $\lambda$, $\lambda_\Sigma $, and $\xi$ are of ${\cal O}(1)$ in the following.
The $\mathbb{Z}_2$ symmetry equates all the dimensionless couplings except the mass parameters in the two sectors: we assume mass hierarchy $M_D\,,M_D'\,,\mu_D \ll M_V\,,M_V'\,,\mu_V$.

\subsubsection{Symmetry Breaking Patterns}

$SU(5)_\VGUT$ is broken down to the gauge group of the SM, $G_\mathrm{SM}$, by a vacuum expectation value (VEV) of $\Sigma_V$ at the scale of $M_\VGUT \simeq \mu_V$, while $X$'s and $H$'s do not obtain large VEVs.
That is, $SU(5)_\VGUT \to G_\mathrm{SM}$ is achieved by
\begin{align}
\langle{\Sigma_V}\rangle = {\cal O}(\mu_V) \, , \quad \langle{X_V\ovl X_V}\rangle = 0 \ .
\end{align}
We set $M_\VGUT = \Ocal(10^{16})\,$GeV, which is expected from the unification of extrapolated gauge coupling constants in the supersymmetric SM (SSM).

The vacuum of dark sector is chosen to be,
\begin{align}
\langle{\Sigma_D}\rangle = {\cal O}(\mu_D)\, , \quad \langle{X_D\ovl X_D }\rangle = {\cal O}(M'_D\MPl) \ .
\end{align}
The non-vanishing VEV of $ {X_D\ovl X_D }$ is due to the forth term of \cref{eq:sp}.
For $\mu_D\sim M'_D \ll \MPl$, $SU(5)_\DGUT$ is first broken down to $SU(4)_{\DGUT}$ by $\langle{X_D}\rangle$.
$SU(4)_{\DGUT}$ is subsequently broken down to $SU(3)_D\times U(1)_D$ by $\langle{\Sigma_D}\rangle$ at $M_\DGUT = {\cal O}(\mu_D)$.
The dark sector results in the model of a composite ADM model in~\cite{Ibe:2018juk,Ibe:2018tex}.

It should be emphasized that the difference between $M_\VGUT$ and $M_\DGUT$ is advantageous to explain the tiny kinetic mixing between the dark photon and the visible photon~\cite{Ibe:2018tex}.
In fact, a higher-dimensional operator,
\eqs{
\label{eq:mixing}
W_\mathrm{Pl} & = \frac{1}{\MPl^2} \tr(\Si_V \mathcal{W}_V) \tr(\Si_D \mathcal{W}_D) \ ,
}
leads to the kinetic mixing parameter of the visible and the dark photons,
\eqs{
\ep \simeq \frac{M_\VGUT M_\DGUT}{\MPl^2} \simeq 10^{-10} \left(\frac{M_\DGUT}{10^{10}~\mathrm{GeV}}\right) \,.
}
We obtain a tiny kinetic mixing parameter $\epsilon = 10^{-7}$--$10^{-10}$ for $M_\DGUT = 10^{10\text{--}13}$\,GeV,
which satisfies all the constraints including
the beam dump experiments~\cite{Bauer:2018onh}
and supernova 1987A~\cite{Chang:2016ntp,Chang:2018rso}
when the dark photon mass is ${\cal O}(10^{1\text{--}2})$\,MeV.

\subsubsection*{Intermediate-Scale Effective Theory}

Below $M_{\VGUT}$, we assume the SSM for the visible sector, where a pair of Higgs doublets from ($H_V$, $\ovl H_V$) remains almost massless by tuning $M_V$ in \cref{eq:sp}.
All the other components of the extra Higgs have masses of ${\cal O}(M_\VGUT)$ in the visible sector.

In the dark sector, $SU(5)_{\DGUT}$ is broken down to $SU(4)_\DGUT$ at $\sqrt{M_\DGUT \MPl}\sim10^{14\mbox{--}16}$\,GeV
for $M_\DGUT = 10^{10\mbox{--}13}$\,GeV.
The gauge multiplets and the pseudo-Goldstone components of $(X_D,\ovl{X}_D)$
corresponding to $SU(5)_\DGUT/SU(4)_\DGUT$
obtain masses of ${\cal O}(\sqrt{M_\DGUT \MPl})$%
\footnote{The mass of the $SU(4)_{\DGUT}$ singlet component of $(X_D,\ovl X_D)$ is of ${\cal O}(M_\DGUT)$. }.
Below the $SU(5)_{\DGUT}$ breaking scale,
the matter and the Higgs multiplets are decomposed into the $SU(4)_\DGUT$ multiplets by
\begin{align}
\Psi_D\to& A_D(\mathbf{6})\oplus Q_D(\mathbf{4})\ ,\,\,
\Phi_D \to \ovl{Q}_D(\mathbf{{\ovl 4}})\oplus N_D(\mathbf{1})\ , \\
H_D\to& H_D(\mathbf{4})
\oplus S_D(\mathbf{1})\ ,
\,\,\ovl H_D\to \ovl H_D(\mathbf{\ovl 4})
\oplus \ovl S_D(\mathbf{1})\ ,
\label{eq:5H}
\\
\Sigma_D\to& \Xi(\mathbf{15}) \oplus h'_D(\mathbf{\ovl 4}) \oplus \ovl h'_D(\mathbf{\ovl 4})
\oplus S'_D(\mathbf{1})\ .
\label{eq:24H}
\end{align}
Below $M_\DGUT$, $SU(4)_\DGUT$ is broken down to $SU(3)_D\times U(1)_D$.
We assume a pair of $U(1)_D$ charged Higgs multiplet remains almost massless
while all the other components in \cref{eq:5H,eq:24H} obtain masses of ${\cal O}(M_\DGUT)$.
The $U(1)_D$ charged Higgs multiplet will break the $U(1)_D$ symmetry at the low energy scale.

Since $(S_D, \ovl S_D)$ do not obtain the VEVs, the matter fields in the dark sector do not obtain
masses from the Yukawa interactions in \cref{eq:sp}.
To generate the mass term, we assume interactions to $X_D$'s such as,
\begin{align}
W = & y_u \Psi_D \Psi_D X_{D} + y_d \Psi_D \Phi_D\ovl X_{D} \cr
& + \frac{y'_e}{\MPl} \Psi_D {\Sigma_D}\Phi_D \ovl X_{D}\ ,
\label{eq:dmass}
\end{align}
with tiny coupling constants%
\footnote{
The corresponding operators in the visible sector do not affect the phenomenology of the SSM.
}.
In the following, we take the masses of the dark quarks to be free parameters.
For a successful model of ADM, the dynamical scale of $SU(3)_D$, $\Lambda_\mathrm{QCD'}$, should be of $\Ocal(1)$\,GeV.
At least, one generation of the quarks should be lighter than $\Lambda_\mathrm{QCD'}$ so that the lightest dark baryon can be the DM%
\footnote{
The dark baryon self interactions mediated by the dark mesons may also realize the velocity dependent cross section, and its implication for structure formation is worth investigating (see Ref.~\cite{Tulin:2017ara} for a review).}.
The last term in \cref{eq:dmass} split the masses of the dark quarks and leptons in $A_D$, $Q_D$, and $\ovl Q_D$.
We assume that the lightest dark lepton is heavier than $\Lambda_\mathrm{QCD'}$ so that
the rapid dark matter decay is avoided~\cite{Ibe:2018tex}.

The visible and dark sectors are connected through superpotential $W_N$ of the right-handed neutrinos.
\eqs{
W_N & = \Phi_V y_N \ovl N H_{V} + \Phi_D y_N \ovl N' H_{D} \\
& ~ + \Phi_V Y_N \ovl N' H_{V} + \Phi_D Y_N \ovl N H_{D} \\
& ~ + \text{(mass terms)} \, ,
\label{eq:wn}
}
where $y_N$ and $Y_N$ are Yukawa coupling constants.
The mass terms of the right-handed neutrinos (denoted by $M_R$ collectively) softly break  $U(1)_X$.
Couplings of $\ovl{N}$  to $\Phi_V$ realize thermal leptogenesis and tiny neutrino masses via the type-I seesaw mechanism~\cite{Minkowski:1977sc,GellMann:1980vs,Preskill:1979zi,Yanagida:1979as,Glashow:1979nm,Mohapatra:1979ia},
while the couplings of $\ovl{N}'$ are irrelevant because we assume that  $\ovl{N}'$ is much heavier than $\ovl{N}$.

The dark neutrinos (included in $\Phi_D$'s) can easily have either Majorana or Dirac mass terms of ${\cal O}(M_{\DGUT})$, and thus our framework is consistent with cosmological constraints on light particles.
For example, the Majorana mass would be generated from $U(1)_X$ breaking higher-dimensional operators such as $(X_D \Phi_D)^2$, while the Dirac mass would be generated from the usual Yukawa coupling, $X_D \Phi_D \ovl N'$.

As shown in Ref.~\cite{Ibe:2018tex}, the $B-L$ portal operators between the two sectors are generated by integrating out the right-handed neutrino and the dark-colored Higgs;
\eqs{
W_\mathrm{eff.} = \frac{(Y_d)_{ij} (Y_N)_{kl}}{\sqrt2 M_C} \ep_{abc} (\ovl U'^a_i \ovl D'^b_j) (\ovl D'^c_k \ovl N_l) \,.
}
Here, $\ovl U'$ and $\ovl D'$ denote the dark quark superfields, and $\ep_{abc}$ is the totally antisymmetric tensor of $SU(3)_D$.
These portal interactions successfully mediate the $B-L$ asymmetry generated by thermal leptogenesis for $M_R < M_C \lesssim 10^2 Y_N Y_d M_R$~\cite{Ibe:2018tex}%
\footnote{For the split supersymmetry with the sfermion masses in ${\cal O}(100)$\,TeV,
the ratio of the VEVs of two Higgs doublets, $\tan\beta$, is required to be $\tan\beta \simeq 2$ to reproduce the observed Higgs boson mass.
In this case, the largest down-type Yukawa coupling is of $\Ocal(10^{-2})$, and hence, a rather large $Y_N$ is required.}.

It should be noted that the above portal interactions require at least two generations of dark quarks to be non-vanishing.
In the following, we leave only the two generations of $U'$ and $D'$ below the $M_\DGUT$ scale, for simplicity.

\begin{figure}[t]
	\includegraphics[width=5.5cm,clip]{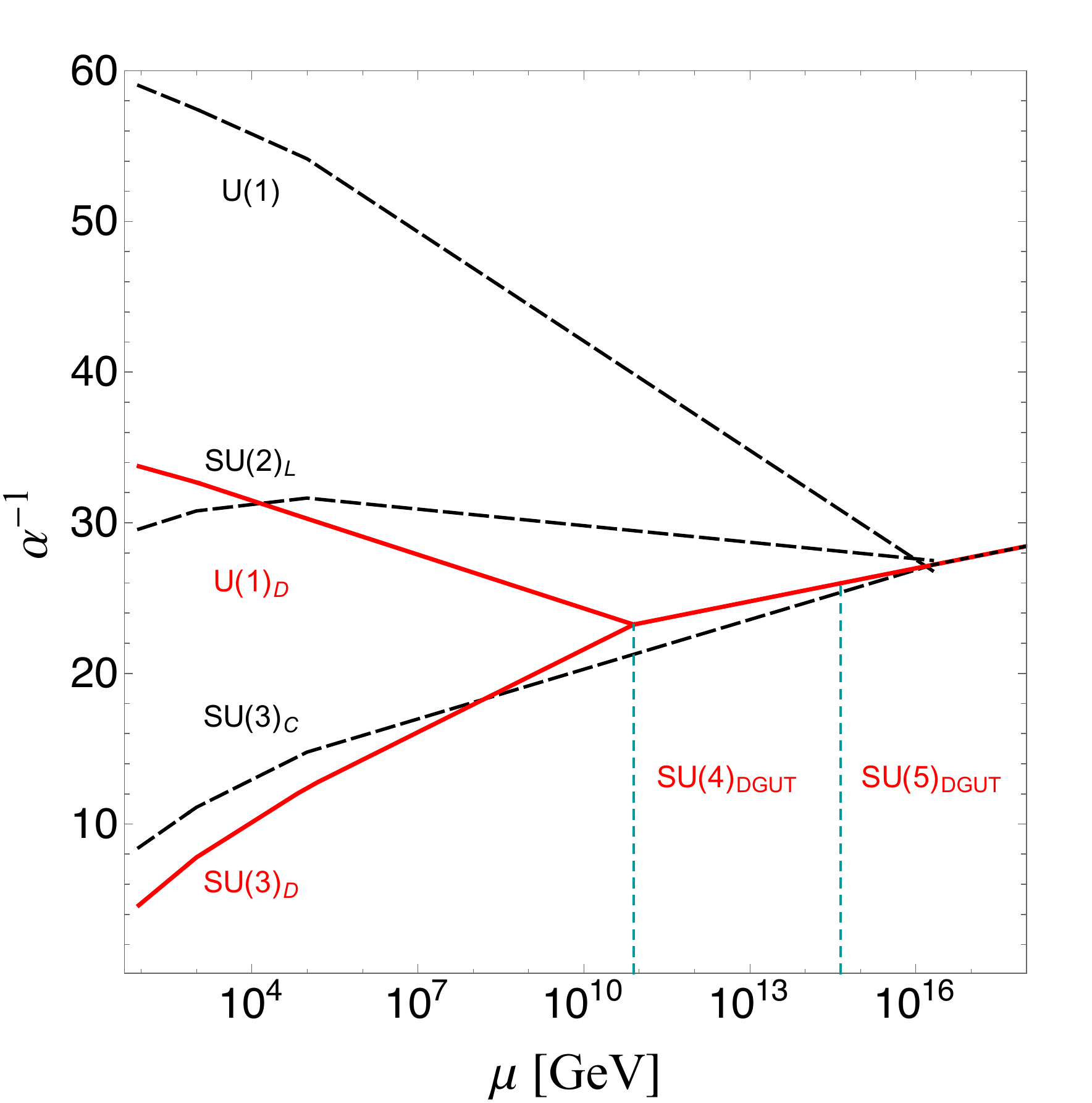}
	\caption{
	Renormalization scale $\mu$ dependence of the gauge couplings in two sectors.
	The broken lines show the (S)SM gauge couplings, while the red lines show the running of dark gauge couplings.
	We assume split spectrum for sparticles: the gaugino scale $1~\mathrm{TeV}$ and the sfermion scale $100~\mathrm{TeV}$.
	We also assume that $M_\DGUT = 8 \times 10^{10}~\mathrm{GeV}$ where $SU(3)_D \times U(1)_D$ unifies into $SU(4)_\DGUT$.
	}
	\label{fig:coupling}
\end{figure}

In \cref{fig:coupling}, we show the one-loop running of the gauge couplings in the two sectors.
We take $M_\DGUT = 8 \times 10^{10}\,\mathrm{GeV}$
and the corresponding  $SU(5)_\DGUT$ breaking scale at $10^{14}$\,GeV as an example.
$M_\DGUT$ of $\Ocal(10^{10})\,\mathrm{GeV}$ or larger is compatible with the composite
ADM scenario with $M_R \gtrsim 10^9\,\mathrm{GeV}$ for thermal leptogenesis~\cite{Ibe:2018tex}.
In this plot, the dark confinement scale $\Lambda_\mathrm{QCD'}$, where $SU(3)_D$ coupling $\alpha_s^{\prime -1}(\Lambda_\mathrm{QCD'})$ vanishes, is about 2.8~GeV which is consistent with the dark baryon mass $m_\mathrm{DM} = {\cal}O(1)\,\mathrm{GeV}$ determined by the asymmetries in two sectors~\cite{Ibe:2011hq,Fukuda:2014xqa,Ibe:2018juk}.
Therefore, the dark baryons with the mass of $\Ocal(1)\,\mathrm{GeV}$ can be naturally realized as a consequence of the $\mathbb{Z}_2$ symmetry at the high-energy scale.

\subsection{Baryon-DM Coincidence}

The dark confinement scale is restricted in our model since the unified couplings in  the two sectors are identified at the GUT scale.
The analytic solution of renormalization group equations for gauge couplings gives the dark confinement scale
\eqs{
\Lambda_\mathrm{QCD'} \simeq 2.8\,\mathrm{GeV} \left( \frac{M_{\mathrm{SUSY}}}{10^2\,\mathrm{TeV}} \right)^{\frac{4}{25}} \left( \frac{M_\DGUT}{8\times 10^{10}\,\mathrm{GeV}} \right)^{\frac{9}{25}} \,,
}
where $M_{\mathrm{SUSY}}$ is a typical mass scale of (dark) sfermions
for two-generation matter in the dark sector below $M_{\DGUT}$.

\begin{figure}[t]
	\includegraphics[width=6cm,clip]{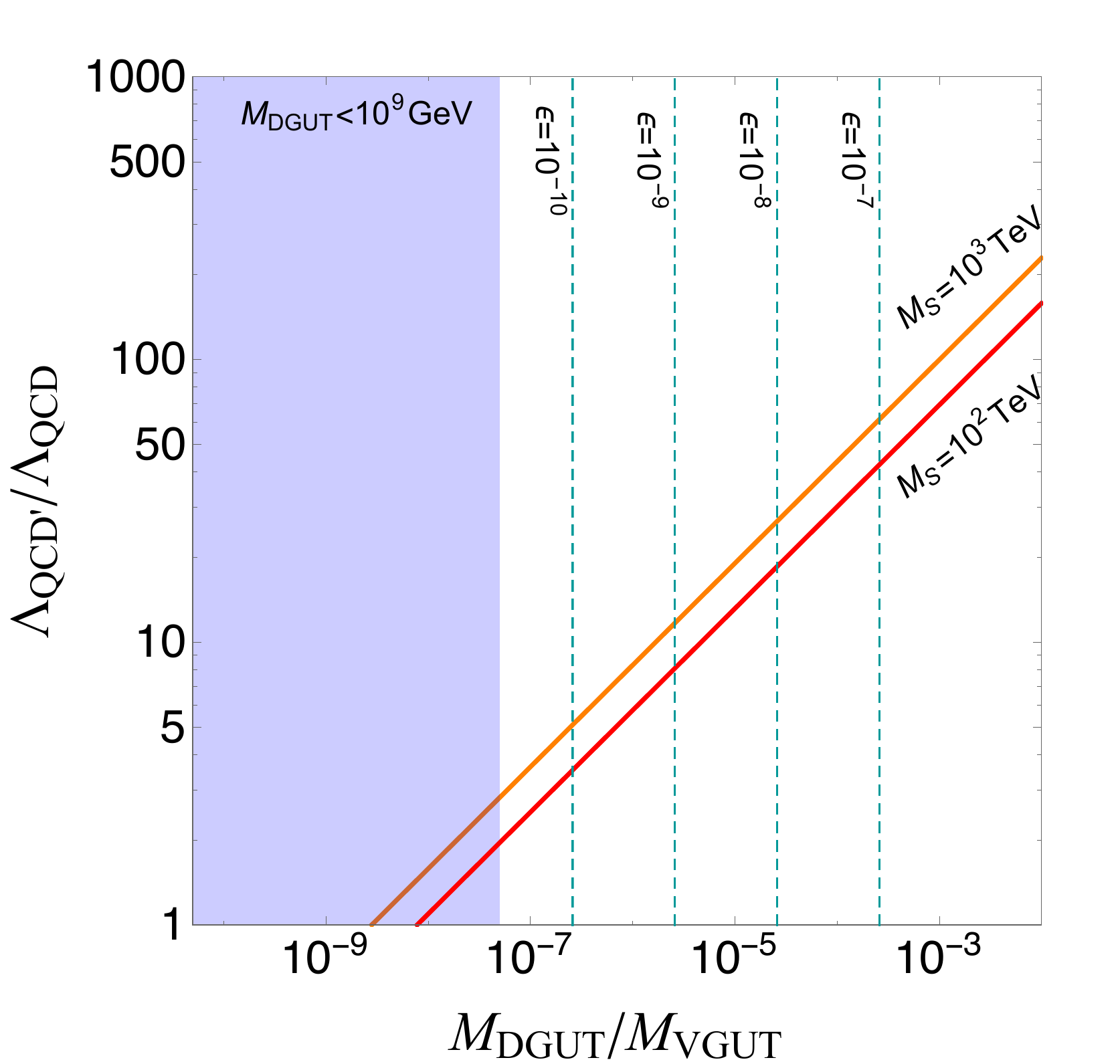}
	\caption{
	Confinement scale as a function of $M_\DGUT/M_\VGUT$.
	Green-dashed lines represent the magnitude of the kinetic mixing parameter $\ep$ given in \cref{eq:mixing}.
	In the blue-shaded region, the ADM scenario which requires $M_{R} < M_{C}\sim M_{\DGUT}$
	is not compatible with the successful thermal leptogenesis which requires $M_R \gtrsim 10^9$\,GeV.
	}
	\label{fig:confinementscale}
\end{figure}

\cref{fig:confinementscale} shows the ratio of the confiment scales in the two sectors.
Here we take $SU(5)_\DGUT$ breaking scale smaller than $M_\VGUT$.
We assume the gauginos and higgsino to be $1\,\mathrm{TeV}$ and the sfermion masses to be $M_{\mathrm{SUSY}} = 10^2\,\mathrm{TeV}~(10^3\,\mathrm{TeV})$ on the red (orange) line.
As a prominent feature of the model, the dark confinement scale is no longer a free parameter in our scenario
and is predicted to be in the range of $\Ocal(1 \text{--} 10^2) \Lambda_\mathrm{QCD}$ for a wide range of $M_\DGUT/M_\VGUT$.
Here we take $\Lambda_\mathrm{QCD} \simeq 0.3\,\mathrm{GeV}$.
This shows that the $\mathbb Z_2$ symmetry 
successfully predicts the dynamical scales are close with each other,
despite the vacuum structures are completely different
between two sectors below the $M_\VGUT$ scale.

It should be also noted that the kinetic mixing parameter is predicted to be $\ep \simeq 10^{-10}$--$10^{-8}$ for $\Lambda_{\mathrm{QCD}' }/\Lambda_\mathrm{QCD} \simeq 5$--$50$.
This feature is another advantage of the present model.

\section{Concluding Remarks}

In this letter, we have proposed the mirrored GUT framework in which the baryon-DM coincidence is naturally explained.
The framework relates the masses of baryon and DM (dynamical scales) and also the number (asymmetry) densities.

In contrast to the models keeping mirror symmetry at a low-energy scale, it is interesting that our framework leads to rich phenomenology and testable signatures~\cite{Ibe:2018juk}.
DM decays into SM neutrinos through the intermediate-energy portal interactions~\cite{Feldstein:2010xe,Fukuda:2014xqa}.
DM annihilates through a dark neutron-antineutron oscillation~\cite{Buckley:2011ye}.
When DM is composed of dark ``charged'' baryons, DM interacts with the SM fermions through tiny kinetic mixing between photon and dark photon.
The monopoles from the $SU(4)_\DGUT \to SU(3)_D \times U(1)_D$ breaking , which are finally confined by the cosmic string after the $U(1)_D$ breaking and annihilate efficiently,  are also worthy of investigation.

We regard the specific model in this paper as a proof of concept and have not addressed the origins of several fine-tuned parameters.
Fine-tunings of parameters are just technically natural thanks to SUSY and furthermore most of tuned parameters are irrelevant to explain the baryon-DM coincidence puzzle.
However, it is to be addressed in future why chiral symmetry breaking in the dark sector is so tiny, although the dark sector is a vector-like theory below the $SU(5)_\DGUT \to SU(4)_\DGUT$ breaking.
We may consider a variant of the present model to ameliorate the parameter tunings in the superpotentials (for example, introducing chiral symmetry to suppress the Higgs $\mu$-term).
Although our present model is not fully satisfactory, it demonstrates a new vast field of the DM-model building to be explored.

\section*{Acknowledgements}
A.~K. would like to acknowledge the Mainz Institute for Theoretical Physics (MITP) of the Cluster of Excellence PRISMA+ (Project ID 39083149) for enabling A.~K. to complete a significant portion of this work.
The work of A.~K. and T.~K. is supported by IBS under the project code, IBS-R018-D1.
This work is also supported by JSPS KAKENHI Grant Numbers, JP17H02878, No. 15H05889 and No. 16H03991 (M.~I.) and by World Premier International Research Center Initiative (WPI Initiative), MEXT, Japan.

\bibliography{ref}

\end{document}